\documentclass[a4paper]{article}
\usepackage{amsmath}
\usepackage{amsfonts}
\usepackage{amsthm,amssymb,upref,amscd}  
\usepackage{ae}
\usepackage{aecompl}

\newcommand{\art}[6]{{\sc #1, \rm #2, \it #3 \bf #4 \rm (#5), \mbox{#6}.}}

\newcommand{\book}[3]{{\sc #1, \it #2, \rm #3.}}
\newcommand{\AND}{{\rm and }}

\title{Relating the Newman--Penrose constants to the Geroch--Hansen multipole moments}
\author{Thomas B\"ackdahl\\
\small{School of Mathematical Sciences,
Queen Mary, University of London,}\\
\small{Mile End Road, London E1 4NS, England}\\
\small{t.backdahl@qmul.ac.uk}}
\date{}

\begin{document}

\maketitle
\begin{abstract}
In this paper, we express the Newman--Penrose constants in terms of the Geroch--Hansen multipole moments for stationary spacetimes. 
These expressions are translation-invariant combinations of the multipole moments up to quadrupole order, which do not normally vanish.
\end{abstract}

\section{Introduction}
The Newman--Penrose (NP) constants were defined by Newman and Penrose in \cite{NewmanPenrose}. They are quantities defined on the null-infinities, and turn out to be conserved under time translations. 
Even though they have been studied for a long time, their meaning is still not fully understood. Lately, it has been disputed whether the NP constants are zero for stationary spacetimes or not. For the Kerr solution they are zero \cite{bai}. In fact, it has been shown that they are zero for all algebraically special stationary spacetimes~\cite{WuShang}. The NP constants have also been calculated for a wide set of examples \cite{DainValiente, FriedrichKannar, LazkozValiente}. The original paper \cite{NewmanPenrose} by Newman and Penrose gives expressions of the NP constants in terms of multipole moments. It is unclear, however, how these moments were defined, if they are coordinate independent and if different moments can be specified independently. The Geroch--Hansen multipole moments have these properties, but were defined later~\cite{Geroch,Hansen}. 
Therefore, this paper is intended to clearly settle the matter by expressing the NP constants in terms of the Geroch--Hansen multipole moments. These multipole moments also give a possibility of physical interpretation. 

The Geroch--Hansen multipole moments can be freely specified under a simple convergence condition. That is, for any given choice of multipoles, satisfying the convergence condition, there is a unique stationay spacetime with these multipole moments. This was shown in \cite{backdahl} for the stationary axisymmetric case. Recently, Herberthson~\cite{herberthson} showed this for the general static case using results of Friedrich~\cite{Friedrich}. The result of Frierdich states that for static spacetimes one can freely specify null data under a convergence condition. These null data are related to the multipole moments, but the relation is fairly complicated. The results by Friedrich have been extended to the stationary case by Ace\~na~\cite{Acena}. Hopefully, the results by Herberthson can also be extended to the stationary case, but for now it is still an open problem. 

For the static case, one could establish the relation between the NP constants and the Geroch--Hansen multipole moments, using the results of Friedrich and K\'ann\'ar~\cite{FriedrichKannar}, but it will not give the general stationary case. One would also need to be careful with the translation between formalisms. Therefore, the original definition of the multipole moments, and the asymptotic expansions of Wu and Shang~\cite{WuShang} are used in this paper.

Throughout this paper we use abstract index notation. 
For coordinate expressions we sometimes omit the indices, 
and use the short hand notation $dx dy = (dx)_{(a} (dy)_{b)}$.

\section{Tetrad expressions}
In this paper, we will use series expansions of stationary spacetimes in Bondi--Sachs coordinates $(u,r,\zeta,\bar\zeta)$. Expressed in standard angular coordinates, the complex angle $\zeta=e^{i\phi}\cot{\frac{\theta}{2}}$. The differential operators $\eth,\bar\eth$ are defined as in equation (4.15.117) in \cite{PenroseRindler1}, for the complex stereographic coordinates $\zeta,\bar\zeta$, i.e. 
\begin{align}
\eth f &= \frac{1+\zeta\bar\zeta}{\sqrt{2}}\frac{\partial f}{\partial\bar\zeta}+ s\frac{\zeta}{\sqrt{2}}f ,&
\bar\eth f &= \frac{1+\zeta\bar\zeta}{\sqrt{2}}\frac{\partial f}{\partial\zeta}- s\frac{\bar\zeta}{\sqrt{2}}f
\end{align}
where $s$ is the spin-weight of $f$. Observe that this differs slightly from the operator usually used for the $\theta,\phi$ coordinates, due to a different choice of spin-frame. 
The corresponding spin-weighted spherical harmonics are then given by
\begin{equation}\label{harmdef}
\begin{split}
{}_sY_{j,m}{}={}& \frac{\sqrt{(2j+1)(j+s)!(j-s)!(j+m)!(j-m)!}
\bar\zeta^{j-m}\zeta^{j+s}}{(-1)^m2\sqrt{\pi}(1+\zeta\bar\zeta)^j}\\
&\times\sum_{r=\max(0,s-m)}^{\min(j-m,j+s)}
{\frac{(-\zeta\bar\zeta)^{-r}}{r!(j-m-r)!(j+s-r)!(r+m-s)!}}
\end{split}
\end{equation}
where $-j\leq s\leq j$, $-j\leq m\leq j$. 

We take the following expansion of the null tetrad from~\cite{WuShang}, using $\Psi_2^0=\bar\Psi_2^0$. 
\begin{equation}
\begin{split}
l^a =&{} \frac{\partial}{\partial r} , \\
n^a =&{} \frac{\partial}{\partial u}+\biggl ( -\frac{1}{2}-\frac{\Psi_2^0}{r}+\frac{\bar\eth\Psi_1^0+\eth\bar\Psi_1^0}{6r^2}-\frac{\bar\eth^2\Psi_0^0+\eth^2\bar\Psi_0^0}{24r^3} \\
&-\Bigl(\frac{|\Psi_1^0|^2}{12}+\frac{\bar\eth^2\Psi_0^1+\eth^2\bar\Psi_0^1}{120}\Bigr ) r^{-4}+\mathcal{O}(r^{-5}) \biggr )\frac{\partial}{\partial r} \\
&+\biggl ( \frac{1+\zeta\bar\zeta}{6\sqrt{2}r^3}\Psi_1^0-\frac{1+\zeta\bar\zeta}{12\sqrt{2}r^4}\bar\eth\Psi_0^0+\mathcal{O}(r^{-5}) \biggr )\frac{\partial}{\partial \zeta} \\
&+\biggl ( \frac{1+\zeta\bar\zeta}{6\sqrt{2}r^3}\bar\Psi_1^0-\frac{1+\zeta\bar\zeta}{12\sqrt{2}r^4}\eth\bar\Psi_0^0+\mathcal{O}(r^{-5}) \biggr )\frac{\partial}{\partial \bar\zeta} ,\\
m^a =&{} \biggl ( -\frac{\Psi_1^0}{2r^2}+\frac{\bar\eth\Psi_0^0}{6r^3}+\frac{\bar\eth\Psi_0^1}{24r^4}+\mathcal{O}(r^{-5}) \biggr )\frac{\partial}{\partial r} \\
&+\biggl ( \frac{1+\zeta\bar\zeta}{6\sqrt{2}r^4}\Psi_0^0+\mathcal{O}(r^{-5}) \biggr )\frac{\partial}{\partial\zeta}+\biggl ( \frac{1+\zeta\bar\zeta}{\sqrt{2}r}+\mathcal{O}(r^{-5}) \biggr )\frac{\partial}{\partial\bar\zeta} ,
\end{split}
\end{equation}
where the expansions of the Weyl curvature are
\begin{align}
\Psi_0 &= \frac{\Psi_0^0}{r^5}+\frac{\Psi_0^1}{r^6}
+\mathcal{O}(r^{-7}),& \Psi_3 &= \frac{\Psi_3^2}{r^4}
+\frac{\Psi_3^3}{r^5}+\frac{\Psi_3^4}{r^6}+\mathcal{O}(r^{-7}),\nonumber \\
\Psi_1 &= \frac{\Psi_1^0}{r^4}+\frac{\Psi_1^1}{r^5}
+\frac{\Psi_1^2}{r^6}+\mathcal{O}(r^{-7}), 
& \Psi_4 &= \frac{\Psi_4^4}{r^5}+\frac{\Psi_4^5}{r^6}
+\mathcal{O}(r^{-7}),\\
\Psi_2 &= \frac{\Psi_2^0}{r^3}+\frac{\Psi_2^1}{r^4}
+\frac{\Psi_2^2}{r^5}+\frac{\Psi_2^3}{r^6}+\mathcal{O}(r^{-7}).\nonumber
\end{align}

We find that for stationary spacetimes, the timelike Killing vector field, can be expressed as
$t^a= T l^a+n^a+\bar A m^a+ A \bar m^a$, 
where $T$ and $A$ were computed in~\cite{WuShang} from the Killing equations, and found to be
\begin{align}
T &= \frac{1}{2}+\frac{\Psi_2^0}{r}-\frac{\bar\eth\Psi_1^0+\eth\bar\Psi_1^0}{6r^2}+\frac{\bar\eth^2\Psi_0^0+\eth^2\bar\Psi_0^0}{24r^3}+\frac{\bar\eth^2\Psi_0^1+\eth^2\bar\Psi_0^1}{120r^4}-\frac{|\Psi_1^0|^2}{12r^4}+\mathcal{O}(r^{-5}) , \nonumber \\
A &= -\frac{\Psi_1^0}{6r^2}+\frac{\bar\eth\Psi_0^0}{12r^3}+\frac{\bar\eth\Psi_0^1}{40r^4}+\mathcal{O}(r^{-5}) .
\end{align}

\section{The metric and quotient metric}
Expressed in terms of the coordinate basis, the Killing vector is
\begin{equation}
t^a = \frac{\partial}{\partial u} + \mathcal{O}(r^{-5})\frac{\partial}{\partial r}+\mathcal{O}(r^{-5})\frac{\partial}{\partial \zeta}+ \mathcal{O}(r^{-5})\frac{\partial}{\partial \bar\zeta}.
\end{equation}

For further calculations, we need expansions of the metric components. The contravariant metric is given by $g^{ab}= 2l^{(a}n^{b)}-2m^{(a}\bar m^{b)}$. Matrix inversion then gives the covariant metric
\begin{align}
g_{ab} ={}& \bigl(1+2\Psi_2^0 r^{-1}-\tfrac{1}{3}(\bar\eth\Psi_1^0+\eth\bar\Psi_1^0)r^{-2}+\tfrac{1}{12}(\bar\eth^2\Psi_0^0+\eth^2\bar\Psi_0^0)r^{-3}+\mathcal{O}(r^{-4})\bigr)du^2\nonumber\\
&+\biggl(\frac{4\sqrt{2}\bar\Psi_1^0}{3(1+\zeta\bar\zeta)}r^{-1}-\frac{\eth\bar\Psi_0^0}{\sqrt{2}(1+\zeta\bar\zeta)}r^{-2}+\mathcal{O}(r^{-3})\biggr)du d\zeta\nonumber\\
&+\biggl(\frac{4\sqrt{2}\Psi_1^0}{3(1+\zeta\bar\zeta)}r^{-1}-\frac{\bar\eth\Psi_0^0}{\sqrt{2}(1+\zeta\bar\zeta)}r^{-2}+\mathcal{O}(r^{-3})\biggr)du d\bar\zeta+2dudr\\
&+\biggl(\frac{2\bar\Psi_0^0}{3(1+\zeta\bar\zeta)^2}r^{-1}+\mathcal{O}(r^{-2})\biggr)d\zeta^2+\biggl(\frac{2\Psi_0^0}{3(1+\zeta\bar\zeta)^2}r^{-1}+\mathcal{O}(r^{-2})\biggr)d\bar\zeta^2\nonumber\\
&+\biggl(-\frac{4}{(1+\zeta\bar\zeta)^2}r^2+\mathcal{O}(r^{-2})\biggr)d\zeta d\bar\zeta .\nonumber
\end{align}
The norm $\lambda= t^a t_a=2T-2A\bar A$ is
\begin{equation}
\lambda= 1+2\Psi_2^0r^{-1}-\tfrac{1}{3}(\bar\eth\Psi_1^0+\eth\bar\Psi_1^0)r^{-2}+\tfrac{1}{12}(\bar\eth^2\Psi_0^0+\eth^2\bar\Psi_0^0)r^{-3}+\mathcal{O}(r^{-4})
\end{equation}
Furthermore, the twist $\omega_a=-\varepsilon_{abcd}t^b\nabla^c t^d$ has a potential $\omega$, which is defined via $\nabla_a\omega=\omega_a$ and $\omega \rightarrow 0$ as $r \rightarrow \infty$. Observe that the sign convention alternates throughout the literature. A change of the sign corresponds to complex conjugation of the multipole moments.
From the metric we compute
\begin{equation}
(\tfrac{\partial}{\partial r})^a\omega_a=\tfrac{2i}{3}(\eth\bar\Psi_1^0-\bar\eth\Psi_1^0)r^{-3}-\tfrac{i}{4}(\eth^2\bar\Psi_0^0-\bar\eth^2\Psi_0^0)r^{-4}+\mathcal{O}(r^{-5}) .
\end{equation}
An integration then yields
\begin{equation}
\omega=-\tfrac{i}{3}(\eth\bar\Psi_1^0-\bar\eth\Psi_1^0)r^{-2}+\tfrac{i}{12}(\eth^2\bar\Psi_0^0-\bar\eth^2\Psi_0^0)r^{-3}+\mathcal{O}(r^{-4}) .
\end{equation}
The equations for the other components are then satisfied due to the vacuum field equations.

Now consider a conformal compactification $V$ of the 3-manifold of trajectories of $t^a$ with metric
$h_{ab}= \Omega^2(-\lambda g_{ab}+t_at_b)$. We want to choose $\Omega$ such that we can add a point $\Lambda$ (the infinity point) such that $h_{ab}$ extends smoothly to $\Lambda$.
We also demand
\begin{equation}\label{OmegaConditions}
\Omega=0,\quad D_a\Omega=0,\quad D_aD_b\Omega=2h_{ab} \quad \text{at } \Lambda ,
\end{equation}
where $D_a$ is the covariant derivative on $h_{ab}$.
The following choice of conformal factor turns out to be adequate:
\begin{equation}
\Omega=(r^{-1}-\Psi_2^0 r^{-2}+\tfrac{11}{8}(\Psi_2^0)^2r^{-3})^2 .
\end{equation}
The coefficients are chosen so as to make the limit of the Ricci tensor of $h_{ab}$ to vanish.

The coordinates $r,\zeta,\bar\zeta$ will naturally induce coordinates on $V$. With a slight abuse of notation we will use the same name for the induced coordinates. Note that $r$ will be a radial coordinate on $V$ for large $r$.
Unfortunately, the components of the metric $h_{ab}$ will not extend smoothly to $\Lambda$ in the Cartesian coordinates corresponding to the coordinates $(R=r^{-1},\zeta,\bar\zeta)$. Therefore, we need better coordinates to verify that our choice of conformal factor is good.\footnote{For the computation of the multipole moments, we actually do not need better coordinates, but to verify smoothness, we do.} One way to find good coordinates is to compute harmonic coordinates. Hence, we will use asymptotically Euclidian harmonic coordinates $(x,y,z)$. For computational purposes, we also use the corresponding spherical coordinates with complex stereographic angles. Thus,
\begin{equation}
x=\rho\frac{\eta+\bar\eta}{1+\eta\bar\eta},\quad y=-i\rho\frac{\eta-\bar\eta}{1+\eta\bar\eta},\quad z=\rho\frac{\eta\bar\eta-1}{1+\eta\bar\eta}.
\end{equation}
A fairly straightforward computation gives us the new coordinates expressed in terms of the old ones: 
\begin{align}
\rho &= r^{-1}-\Psi_2^0 r^{-2}+\tfrac{5}{4}(\Psi_2^0)^2r^{-3}+\mathcal{O}(r^{-4}) , \nonumber \\
\eta &= \zeta-\tfrac{\sqrt{2}}{6}(1+\zeta\bar\zeta)\Psi_1^0 r^{-2} +\mathcal{O}(r^{-3}) .
\end{align}

The conformal metric and the conformal factor are then found to be 
\begin{align}
h_{ab}&=dx^2+dy^2+dz^2 +\mathcal{O}(\rho^3) ,&
\Omega=&\rho^2+\tfrac{1}{4}(\Psi_2^0)^2\rho^4+\mathcal{O}(\rho^5) .
\end{align}

Now we easily see that $\Omega\rightarrow 0$ when $\rho\rightarrow 0$; thus, $\rho=0$ will now represent the infinity $\Lambda$ on our 3-manifold. The smoothness of $h_{ab}$ and the conditions \eqref{OmegaConditions} can now be easily verified. 
The Ricci tensor $R_{ab}$ of $h_{ab}$ is $R_{ab}=\mathcal{O}(\rho)$.

\section{Geroch--Hansen multipole moments}
Define the complex potential
\begin{equation}
P=\frac{1-\lambda-i\omega}{(1+\lambda+i\omega)\sqrt{\Omega}}.
\end{equation}
This potential as well as the choice of sign in the definition of the twist is taken from \cite{FHP}.
There are many different possible choices of potential, but large classes of potentials do produce the same moments~\cite{SimonBeig}. 
The Geroch-Hansen multipole moments \cite{Geroch,Hansen} are given by the limits of
\begin{equation}
P_{a_1\dots a_n}=C\biggl[D_{a_1}P_{a_2\dots a_n}-\frac{(n-1)(2n-3)}{2}R_{a_1a_2}P_{a_3\dots a_n}\biggr],
\end{equation}
as one approaches $\Lambda$. Here $C[\cdot]$ represents the totally symmetric and trace-free part.

Hence, with monopole (mass) $M$, dipole $C_a$, and quadrupole $Q_{ab}$ expressed in Cartesian coordinates, we by definition have
\begin{equation}\label{momentsMCQ}
\begin{split}
\lim_{\rho\rightarrow 0} P\phantom{{}_{ab}} ={}& M\\
\lim_{\rho\rightarrow 0} P_a\phantom{{}_{b}} ={}& C_x dx+C_y dy+C_z dz\\
\lim_{\rho\rightarrow 0} P_{ab} ={}& Q_{xx} dx^2+Q_{yy} dy^2-(Q_{xx}+Q_{yy})dz^2\\
&+2Q_{xy}dxdy+2Q_{xz}dxdz+2Q_{yz}dydz .
\end{split}
\end{equation}

Under a translation $\Omega'=\Omega(1+xT_x+yT_y+zT_z)$ the dipole will transform like $C'_j=C_j-\tfrac{1}{2} M T_j$, while the quadrupole will transform like
\begin{equation}\label{Qtransform}
\begin{split}
Q'_{xx}&=Q_{xx}-2 T_x C_x+T_y C_y+T_z C_z-\tfrac{1}{4} M\left(-2 {T_x}^{2}+{T_y}^{2}+{T_z}^{2} \right),\\
Q'_{yy}&=Q_{yy}+T_x C_x-2 T_y C_y+T_z C_z-\tfrac{1}{4} M\left({T_x}^{2}-2 {T_y}^{2} +{T_z}^{2} \right) , \\
Q'_{xy}&=Q_{xy}-\tfrac{3}{2} T_x C_y-\tfrac{3}{2} T_y C_x+\tfrac{3}{4} M T_x T_y,\\
Q'_{xz}&=Q_{xz}-\tfrac{3}{2} T_x C_z-\tfrac{3}{2} T_z C_x+\tfrac{3}{4} M T_x T_z, \\
Q'_{yz}&=Q_{yz}-\tfrac{3}{2} T_y C_z-\tfrac{3}{2} T_z C_y+\tfrac{3}{4} M T_y T_z.
\end{split}
\end{equation}

We expand $\Psi_0^0$, $\Psi_1^0$ and $\Psi_2^0$ in terms of spin-weighted spherical harmonics:
\begin{equation}
\begin{split}
\Psi_0^0&=\sum_{m=-2}^2 A_m {}_2Y_{2,m}= \sqrt{5}\frac{A_{-2}+2\zeta A_{-1}+\sqrt{6}\zeta^2 A_0+2\zeta^3 A_1+\zeta^4A_2}{2\sqrt{\pi}(1+\zeta\bar\zeta)^2} , \\
\Psi_1^0&=\sum_{m=-1}^1 B_m {}_1Y_{1,m}= -\sqrt{3}\frac{B_{-1}+\sqrt{2}\zeta B_0+\zeta^2 B_1}{2\sqrt{\pi}(1+\zeta\bar\zeta)} ,\\
\Psi_2^0&=C .
\end{split}
\end{equation}
Here $C$ is real, $B_m$ and $A_m$ are complex.

A series expansion of the potential yiels
\begin{equation}
\begin{split}
P={}&-C+\frac{2\bar\eta B_{-1}+\sqrt{2}(\eta\bar\eta-1)B_0-2\eta B_1}{\sqrt{24\pi}(1+\eta\bar\eta)}\rho-\frac{\sqrt{5}(\eta^2\bar\eta^2-4\eta\bar\eta+1)A_0}{4\sqrt{6\pi}(1+\eta\bar\eta)^2}\rho^2\\
&-\frac{\sqrt{5}(\bar\eta^2 A_{-2}+\bar\eta(\eta\bar\eta-1)A_{-1}-\eta(\eta\bar\eta-1)A_1+\eta^2 A_2)}{4\sqrt{\pi}(1+\eta\bar\eta)^2}\rho^2+\frac{3 C^3}{8}\rho^2+{\mathcal{O}}(\rho^3)
\end{split}
\end{equation}

One then easily obtains the multipole moments by changing to Cartesian coordinates and taking limits: 
\begin{equation}\label{momentsABC}
\begin{split}
\lim_{\rho\rightarrow 0} P\phantom{{}_{ab}}={}& -C , \\
\lim_{\rho\rightarrow 0} P_a\phantom{{}_{b}} ={}& \lim_{\rho\rightarrow 0} D_a P= \tfrac{\sqrt{6}}{12\sqrt{\pi}}(B_{-1}-B_1)dx-\tfrac{i\sqrt{6}}{12\sqrt{\pi}}(B_{-1}+B_1)dy+\tfrac{\sqrt{3}}{6\sqrt{\pi}}B_0 dz , \\
\lim_{\rho\rightarrow 0} P_{ab} ={}& \lim_{\rho\rightarrow 0} (D_aD_b P-\tfrac{1}{3}h_{ab}D^cD_c P) = \tfrac{\sqrt{5}}{24\sqrt{\pi}}(\sqrt{6}A_0-3A_2-3A_{-2})dx^2\\
&+\tfrac{\sqrt{5}}{24\sqrt{\pi}}(\sqrt{6}A_0+3A_{2}+3A_{-2})dy^2+\tfrac{i\sqrt{5}}{4\sqrt{\pi}}(-A_2+A_{-2})dx dy\\
&-\tfrac{\sqrt{30}}{12\sqrt{\pi}}A_0dz^2 +\tfrac{\sqrt{5}}{4\sqrt{\pi}}(A_1-A_{-1})dx dz+\tfrac{i\sqrt{5}}{4\sqrt{\pi}}(A_1+A_{-1})dy dz .
\end{split}
\end{equation}

\section{Newman--Penrose constants}
By comparing the limits \eqref{momentsMCQ} and \eqref{momentsABC}, one can conclude that
\begin{align}
A_{-2}&=-2\sqrt{\tfrac{\pi}{5}} (Q_{xx}-Q_{yy}+2 iQ_{xy}) ,& 
B_{-1}&=\sqrt{6\pi} (C_x+iC_y) ,\nonumber \\
A_{-1}&=-4\sqrt{\tfrac{\pi}{5}} (Q_{xz}+iQ_{yz}) ,&
B_{0}&=2 \sqrt{3\pi} C_z ,\nonumber \\
A_{0}&=2\sqrt{\tfrac{6\pi}{5}} (Q_{xx}+Q_{yy}) ,&
B_{1}&=\sqrt{6\pi}(-C_x+iC_y) , \\
A_{1}&=4\sqrt{\tfrac{\pi}{5}} (Q_{xz}-iQ_{yz}) ,&
C&=-M ,\nonumber \\
A_{2}&=2\sqrt{\tfrac{\pi}{5}} (-Q_{xx}+Q_{yy}+2 iQ_{xy}) .\nonumber
\end{align}

The NP constants $\{G_m\}$ can then be computed from
\begin{equation}
G_m=\int_0^{2\pi}\int_0^\pi{\Psi_0^1 \overline{{}_2 Y_{2,m}}\sin\theta d\theta d\phi}=\int_0^{2\pi}\int_0^\pi{(\tfrac{10}{3}\Psi_1^0-5\Psi_2^0\Psi_0^0)\overline{{}_2 Y_{2,m}}\sin\theta d\theta d\phi} .
\end{equation}
Here the spin-weighted spherical harmonics are as in the definition~\eqref{harmdef}. For the integration, the variables are changed to $(\theta,\phi)$ via $\zeta=e^{i\phi}\cot\frac{\theta}{2}$. Observe that we do not change the spin frame to be adapted to the new coordinates.  
Expansions of the integrands can, in principle, be taken from \cite{WuShang} eq $(51)$, but they do use a different spin-frame in that section the paper, hence it is easier to redo the calculations than translating the result.

The integration gives
\begin{align}
G_{-2}&=-2\sqrt{5\pi}(3 C_y^2-3C_x^2+MQ_{xx}-MQ_{yy}+2iMQ_{xy}-6iC_xC_y), \nonumber \\
G_{-1}&=-4\sqrt{5\pi}(iMQ_{yz}-3C_xC_z-3iC_yC_z+MQ_{xz}), \nonumber\\
G_0&=2\sqrt{30\pi}(-C_x^2-C_y^2+2C_z^2+MQ_{xx}+MQ_{yy}), \\
G_1&=-4\sqrt{5\pi}(iMQ_{yz}+3C_xC_z-3iC_yC_z-MQ_{xz}), \nonumber\\
G_2&=-2\sqrt{5\pi}(3 C_y^2-3C_x^2+MQ_{xx}-MQ_{yy}-2iMQ_{xy}+6iC_xC_y). \nonumber
\end{align}

As expected, this is the same form as in the original paper by Newman and Penrose~\cite{NewmanPenrose}, i.e., linear combinations of dipole squared and monopole times quadrupole. 
From the translation rules \eqref{Qtransform}, it is easy to see that the NP constants are invariant under translations. Hence, they are independent of the choice of conformal factor. As the NP constants are expansion coefficients for spin-weighted spherical harmonics, they will depend on the spin-frame though.

For the axisymmetric case, we see that $G_{-2}=G_{-1}=G_1=G_2=0$ and $G_0=2\sqrt{30\pi}(2C_z^2-MQ_{zz})$, where $Q_{zz}=-2Q_{xx}=-2Q_{yy}$ is the $zz$-component of the quadrupole.

We can conclude that the NP constants are, in general, not zero, but for some important solutions they are. For instance, the Kerr solution has $C_z=iMa$, $Q_{zz}=-2Q_{xx}=-2Q_{yy}=-2Ma^2$, and all other components of $C_a$ and $Q_{ab}$ are zero. This yields the well-known fact that all NP constants are zero for the Kerr solution. In fact, they are zero for all stationary, algebraically special solutions \cite{WuShang}.

\section{Acknowledgements}
This work was supported by the Wenner-Gren foundations.
Thanks to Juan A. Valiente Kroon, for helpful discussions. I would also like to thank Lars Anders\-son for asking about the relation between multipole moments and Newman--Penrose constants.

\end{document}